\documentclass[12pt]{iopart}

\usepackage{iopams}
\usepackage{graphicx}
\usepackage{subfigure}
\begin{document}

\title[$\Xi^{0}(1530)$ Production in Heavy-Ion Collisions]{$\Xi^{0}(1530)$ Production in Heavy-Ion Collisions \linebreak and its Implications for $\Delta t_{therm-chem}$}

\author{R. Witt for the STAR Collaboration}

\address{Yale University, 
WNSL--West, 
272 Whitney Ave., 
New Haven, CT, USA}
\ead{richard.witt@yale.edu}
\begin{abstract}
We present the first measurements of the transverse momentum spectra and mid-rapidity yields of the $\Xi^{0}(1530)$ multi-strange baryon resonance, from $\sqrt{\mathrm{s_{NN}}}=200$ GeV Au+Au collisions at RHIC from the STAR experiment.  Ratios of the mid-rapidity yield of this long-lived ($c\tau=21$ fm) resonance to its non-resonance partner (the $\Xi$ baryon) are compared to predictions from thermal models and to the corresponding ratios for other resonances previously reported by STAR.  Conclusions will be drawn about the implication of these measurements for the time between chemical and thermal freeze-out ($\Delta t_{therm-chem}$) and any observed deviations from the expected mass and width will be discussed.
\end{abstract}



\section{Introduction}

Hadronic scattering in the late stages of a heavy-ion collision may modify the chemical composition of the system up to the time of chemical freeze-out, defined as when all non-resonant particle ratios are fixed.  However, resonances may still be produced through inelastic hadronic scatterings, their production rates being dependent on the relatively poorly known hadronic cross-sections ($\Xi^{-}+\pi^{+}\rightarrow\Xi^{*}(1530)$, $\Lambda+\pi^{\pm}\rightarrow\Sigma^{*}(1385)$, etc.) \cite{Bass00,Rapp01,vanHees05,Adler02}.  Similarly, rescattering of the daughters of resonances that decay during the lifetime of the colliding system may destroy their correlations.  This prohibits the parent resonance reconstruction, leading to an underestimated primary yield.  The interplay of these two effects continues from the time of chemical freeze-out ($t_{chem}$) to the time of thermal freeze-out ($t_{therm}$, currently estimated to be $\sim$8-10 fm/$c$ \cite{Retiere04}), the degree to which the primary resonance yields are modified depends on both the length of this time interval ($\Delta t = t_{therm} - t_{chem}$) and the relative strengths of the two effects \cite{Bleicher02, Bleicher03, Torrieri01, Rafelski01, Rafelski02}.  It has also been suggested that the \textit{in vacuo} resonance masses and widths might be modified in the presence of the nuclear medium \cite{STARRes}.

The STAR experiment \cite{STAR} has measured several mesonic and baryonic resonance states \cite{STARRes, STARK*, STARrho, STARphi}.  The variation in the composition and lifetimes of the measured states provides multiple constraints on $\Delta t_{therm-chem}$ and the degree of late stage hadronic interactions.  The measurements presented herein add to the emerging picture of how the medium produced in heavy-ion collisions at RHIC affects resonance production.

\section{Analysis}

The results presented here come from an analysis of 7.6$\times 10^{6}$, 5.4$\times 10^{6}$, and 6.8$\times 10^{6}$ Au+Au events in the 0-12\%, 10-40\%, and 40-80\% centrality ranges respectively.  The greatest difficulty in performing an analysis of short-lived particles is acquiring an accurate description of the underlying combinatorial  background.  For this analysis, we use a narrow rotational method that is implemented as follows.  

\begin{figure}
 \centering
 \includegraphics[bb= 0 0 567 547, scale=0.45]{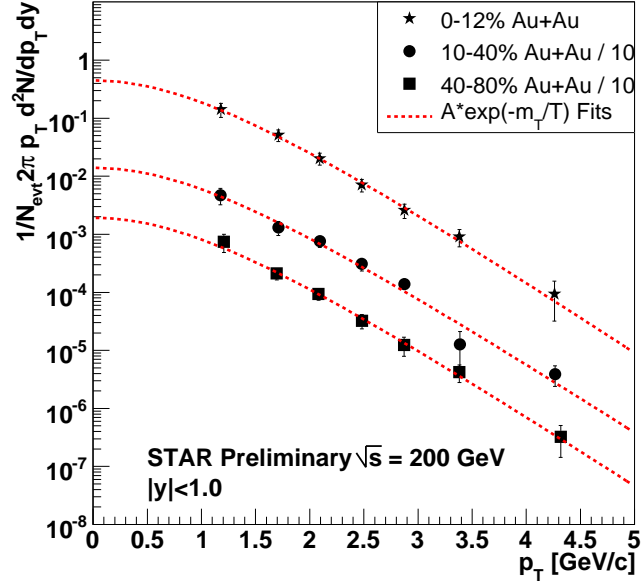}
 \caption{Corrected $\Xi^{0}(1530)$ $\mathrm p_{\mathrm T}$ spectra from Au+Au collisions at $\sqrt{\mathrm{s}_{\mathrm NN}}=200$ GeV/$c$ for three centrality bins.}
 \label{spectra}
\end{figure}

In each event, the invariant mass is calculated for each topologically found $\Xi^{-}$ candidate that passes a set of quality cuts combined with each $\pi^{+}$ candidate that has passed a set of particle identification cuts.  These combinations comprise the ``signal''.  A second set of combinations is then made in which the transverse momentum ($\mathrm p_{\mathrm T}$) vectors of the $\Xi^{-}$ candidates are rotated in the azimuthal plane by angles from 170$^{\circ}$ to 188$^{\circ}$ and the $\pi^{+}$ candidates are rotated from -8$^{\circ}$ to +8$^{\circ}$, both in steps of 2$^{\circ}$.  The invariant mass is then calculated for each iteration and each pair.  These combinations comprise the ``background''.  The process is repeated as a function of $\mathrm p_{\mathrm T}$ and the background is then subtracted from the signal in each bin.  Lastly, the yield is determined by summing the bin contents in a narrow region around the canonical mass.  Efficiency and acceptance corrections are applied to the results.  Lastly, a correction is applied to account for contribution to the overall yield coming from the neutral decay channel, which we do not measure here.  The corrected mid-rapidity spectra for three centralities are show in Figure \ref{spectra}.

\section{Results and Conclusions}

\begin{figure*}[h]
\hspace*{-1.0cm}
\subfigure[Resonance to non-resonance ratios for several species in $p+p$, d+Au, and Au+Au collisions.]{
\hspace*{-1.0cm}
 \label{ratios}
 \includegraphics[bb= 6 5 561 542, scale=0.45]{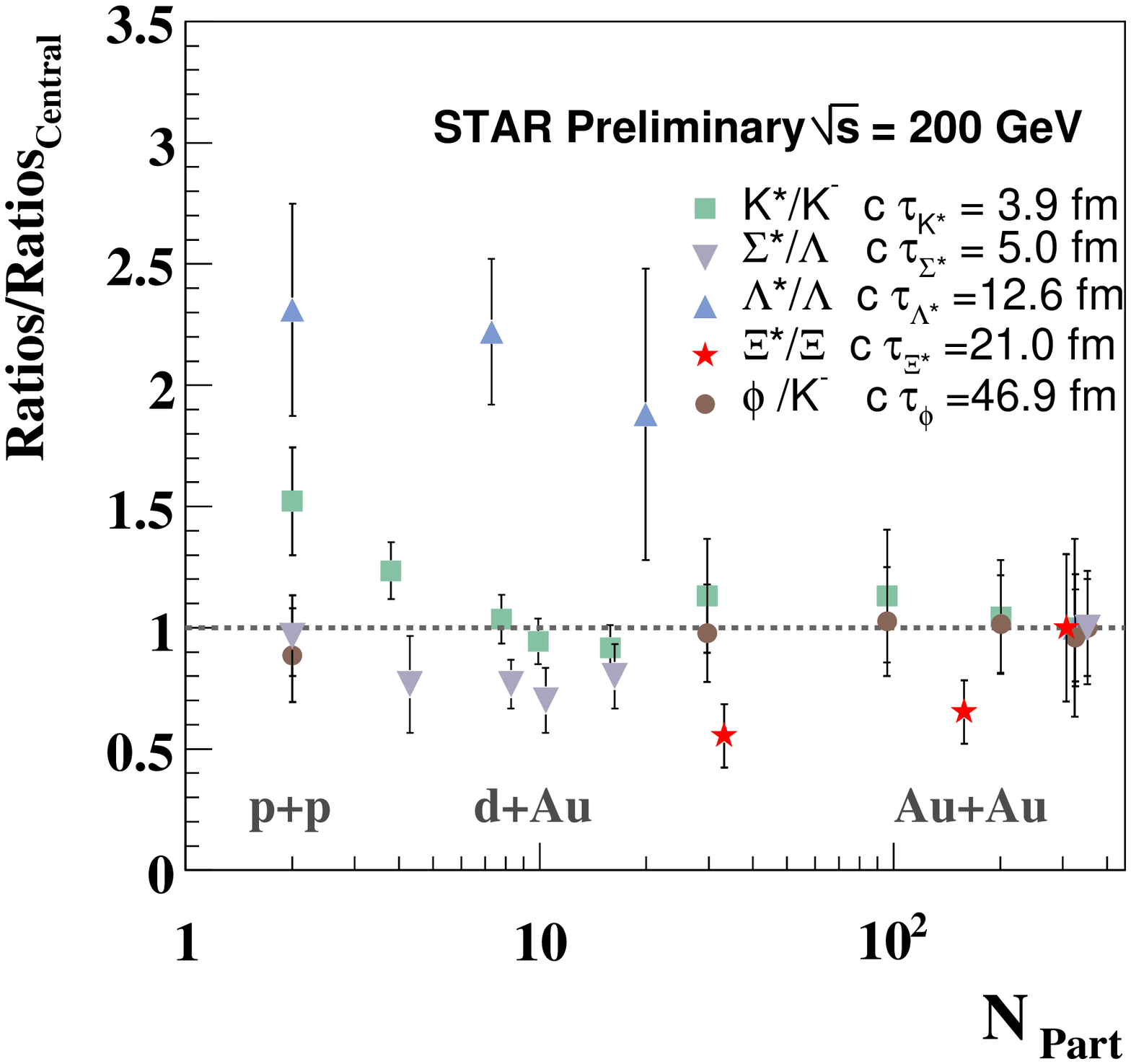}}
\hspace*{0.1cm}
\subfigure[Thermal model calculation.  Black dots are data, model fit and predictions are horizontal lines.  Data to the right side of the ``Fitted'' line were not included in the fit, but rather compared to predictions from the fit.]{
 \label{thermus}
 \includegraphics[bb= 0 0 567 547, scale=0.45]{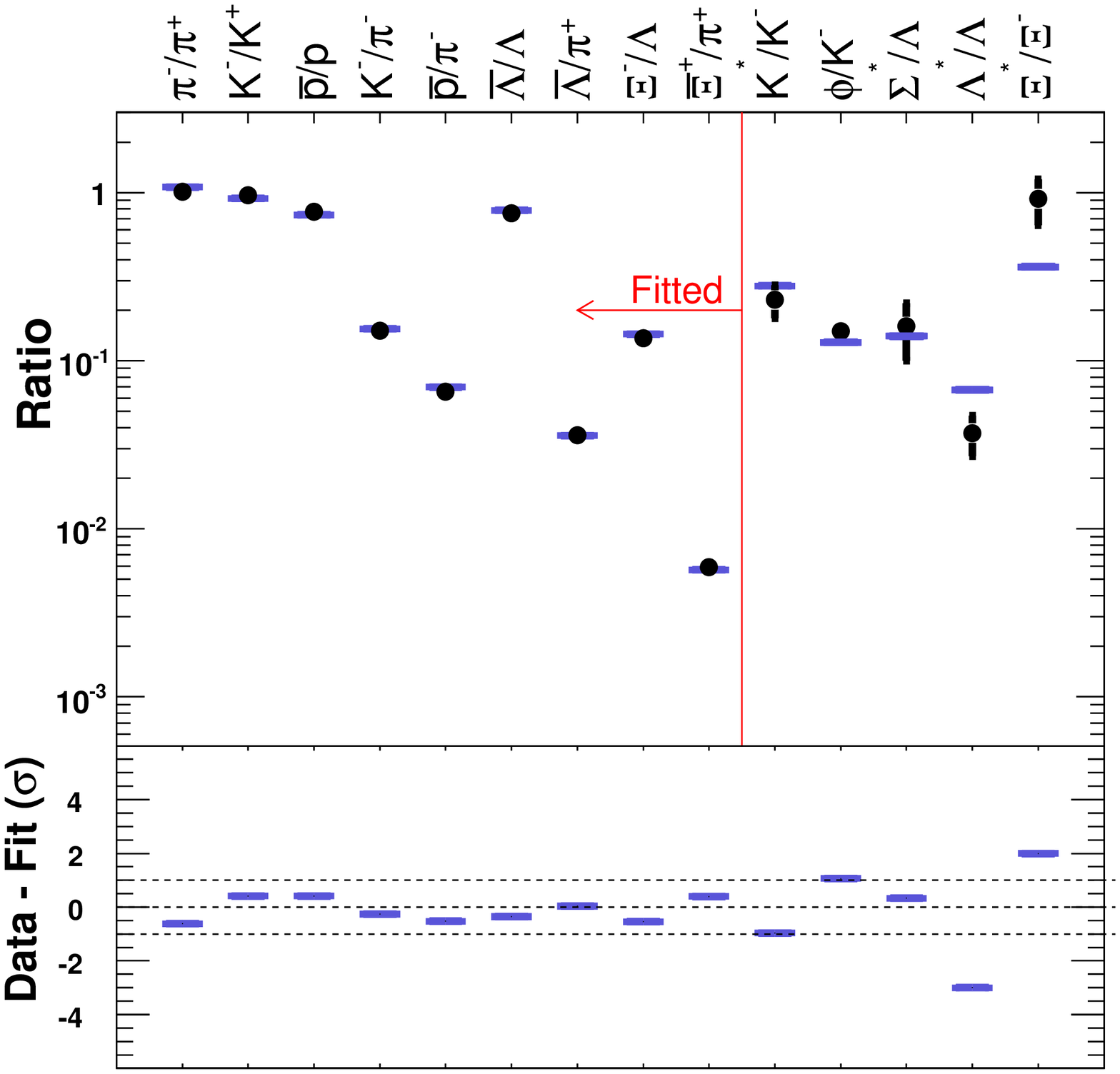}}
\end{figure*}

Information about medium effects on resonance production is gained by comparing the resonance yield to that of its non-resonant partner.  In Figure \ref{ratios} we show this comparison, plotted as a ratio of the resonance to non-resonance yield, for the relevant species measured by STAR.  The mid-rapidity ($|y|<1.0$) $\Xi^{0}(1530)$ $dN/dy$ yields are compared with ($|y|<0.75$) $\Xi$ $dN/dy$ yields from reference \cite{STARXi}.
These comparisons are usually shown scaled by the $p+p$ value.  However, as there is currently no measured $p+p$ ratio for the $\Xi^{0}(1530)$, the ratios have all been scaled by the most central Au+Au values.  This alternative scaling results in what is usually seen as a supression in central Au+Au appearing as an enhancement in peripheral (and $p+p$) collisions and an enhancement in central Au+Au appearing as a suppression in peripheral (and $p+p$) collisions.

Examining the ratios we find the following: The shortest lived resonance measured by STAR, the K$^*$, shows a small suppression in central collisions relative to peripheral \cite{STARK*ratio}, and the only slightly longer-lived $\Sigma^{*}(1385)$ shows no measurable effect \cite{STARRes}.  The $\Xi^{0}(1530)$ ratios show an \textit{enhancement} in the most central collisions relative to the peripheral collisions, the opposite of what is seen for the other reasonably long-lived baryonic resonance, the $\Lambda^{*}(1520)$, which shows a strong suppression in central collisions relative to $p+p$.

The short-lived resonances would be expected to show suppression in more central events due to the rescattering of their decay products.  The fact that they are only weakly suppressed, if at all, suggests there must be significant regeneration contributing to the final state yields.  This conclusion is strengthened when one adds the $\Xi^{0}(1530)$ measurement to the picture.  The $\Xi^{0}(1530)$ should easily outlive the system (on average) once it's produced.  This means rescattering does not significantly affect the $\Xi^{0}(1530)$ yield, (re-)generation plays a stronger role.  The observed $\Xi^{0}(1530)$ enhancement in central collisions seems to indicate that significant interaction, and resonance production, occurs between $\Xi$ and $\pi$ particles in the (late) hadronic phase.  The $\phi$ meson, like the $\Xi^{0}(1530)$, will not on average decay before $t_{therm}$.  However, hadronic phase $\phi$ production may be suppressed by the relatively low probability for K$^{+}$+K$^{-}$ scattering in the final stages of the collision \cite{STARphi}.  The $\Lambda^{*}(1520)$ is also reasonably long-lived.  However, alternative explanations have been put forward for the observed $\Lambda^{*}(1520)$ suppression that do not involve rescattering and regeneration \cite{Enyo,Kaskulov}.

Lastly, in Figure \ref{thermus} we show the results of a thermal model fit to the published STAR data from central $\sqrt{\mathrm{s}_{\mathrm NN}}=200$ Au+Au collisions.  The fit was performed using the THERMUS interface \cite{thermus}.  The resulting fit parameters are T=0.169$\pm$0.006 GeV, $\mu_{\mathrm B}$=0.04$\pm$0.01 GeV, $\mu_{\mathrm S}$=0.016$\pm$0.009 GeV, $\mu_{\mathrm Q}$=-0.01$\pm$0.01 GeV, $\gamma_{\mathrm S}$=0.91$\pm$0.06, and radius=7.5$\pm$1.0 fm.  All of the resonance to non-resonance ratios are shown to the right of the ``Fitted'' line compared to their predicted values from the thermal fit.  The same pattern of suppression and enhancement as was seen in the ratios alone shows-up again.  The thermal model gives predictions for particle ratios only at $t_{chem}$, and therefore does not account for production that occurs during the hadronic phase, and so the observed $\Xi^{0}(1530)$ enhancement could again be attributed to significant resonance production occurring during the hadronic phase.  This result strengthens the conclusion drawn from the resonance to non-resonance ratios alone.

\end{document}